\documentclass[%
 reprint,
 amsmath,amssymb,
 aps,
]{revtex4-1}

\usepackage[dvipdfmx]{graphicx}
\usepackage{dcolumn}
\usepackage{bm}
\usepackage{amsmath}
\usepackage{amssymb}
\usepackage{mathrsfs}
\usepackage{natbib}

\newcommand{\mathsfbi}[1]{\boldsymbol{\mathsf{#1}}}

\begin{document}


\title{Mean-Lagrangian formalism and covariance of fluid turbulence}
\author{Taketo Ariki}
\email{ariki@fluid.cse.nagoya-u.ac.jp}
\affiliation{Institute of Material and Systems for Sustainability, Nagoya University, Furocho, Chikusa-ku, Nagoya, Japan}
\date{\today}

\begin{abstract}
Mean-field-based Lagrangian framework is developed for the fluid turbulence theory. The space-time vector flow is naturally introduced from the mean velocity, which provides the Lagrangian picture based on the mean field in totally equivalent manner to that of the instantaneous field. The proposed framework is applied, on one hand, to one-point closure model, yielding an objective expression of the turbulence visco-elastic effect. Application to two-point closure, on the other hand, is also discussed, where natural extension of known Lagrangian correlation is discovered on the basis of extended covariance group.
\begin{description}
\item[PACS numbers] 
\end{description}
\end{abstract}

\pacs{Valid PACS appear here}
\maketitle

\section{Introduction}

In general continuum mechanics, effective fluxes in continuum scale are caused by the thermodynamic imbalance of microscopic nature within finite relaxation time, where the constitutive relations depend on the history of various physical properties. The convective-coordinate framework, or the \emph{Lagrangian framework}, offers the most suitable and natural description of such history effect \citep{Eringen75}. On the contrary, the classical fluid mechanics, as a prototypical model of the continuum, has enjoyed its field-theoretic development due to simpler coordinate descriptions of the \emph{Eulerian framework}, since the pure viscous fluid, including the Newtonian fluid, is an idealized model of infinitesimal relaxation time, where the history effect does not appear. 

At the stage of turbulent flow, however, the turbulence motion causes finite relaxation time in the mean-field properties, requiring the reconsideration of history effect at the mean-field level. In addition, the covariance principle also survives in turbulence-constitutive relations as a rigorous requirement, which is an essential ingredient for the physical objectivity of the observable mean fields \citep{Ariki15a}. These facts naturally motivate the fundamental reformulation of the turbulence theory in the Lagrangian framework of the mean-field base, which may be referred as the \emph{mean-Lagrangian framework} hereafter. Furthermore, a remarkable similarity between instantaneous- and mean-velocity fields are pointed out by recent \citep{Ariki15a} (termed as the \emph{v-V correspondence}), stating that the mean velocity field plays equivalent roles to those of the instantaneous counterpart, which also supports the possibility of such mean-Lagrangian reformulation. 


In this paper, we clarify that the mean-Lagrangian framework provides a firm mathematical basis suitable to the covariant formulation of the theory of fluid turbulence. Space-time description elucidates the true equivalence between instantaneous- and mean-velocity fields, both of which appear as one-parameter groups (vector flows) of space-time. The proposed framework is applied to one-point closure model, which allows an explicit integration of the visco-elastic effect of the Reynolds stress in generally covariant manner. Application to the two-point closure is also considered, where we propose some new candidates of turbulence correlation allowed by the mean-Lagrangian description, which include some generalizations of the conventional Lagrangian quantities of \citep{Krai65,Kaneda81}.


\section{4-dimensional formalism}
We observe the fluid motion in the space $\mathscr{S}$ as time $t(\in\mathbb{R})$ passes, which may be clearly illustrated by the 4-dimensional space-time formulation. We compose the space-time as the direct product $\mathscr{M}=\mathscr{S}\times\mathbb{R}$, and express the space $\mathscr{S}$ at each time $t$ as a 3-D hyper surface $\mathscr{S}_t$ of $t=const.$. Here we only discuss the non-relativistic limit, where the coordinate transformation does not change the time $t$. Namely the non-relativistic general transformation is given by
\begin{equation}
\tilde{\mathbf{x}}=\tilde{\mathbf{x}}(\mathbf{x},t),\  
\tilde{t}=t.
\label{S transform}
\end{equation}
In the space-time formalism, $t$ may be written as $x^{{}_4}$, and the above transformation forms the subgroup of 4-dimensional diffeomorphism group such that $x^{\tilde{\mu}}=x^{\tilde{\mu}}(x^{{}_1},x^{{}_2},x^{{}_3},x^{{}_4})$. This transformation enables us to identify bodies, including fluid, as hyper-surfaces embedded in the 3-dimensional slice of space-time $\mathscr{M}$. Note that general tensor $T$ in $\mathscr{M}$ does not behave as tensor in $\mathscr{S}_t$; the general 4-D 1-rank tensor $T$ transforms as $T^{\tilde{\alpha}}=x^{\tilde{\alpha}}{}_{,\mu}T^\mu$, whose spatial component becomes
\begin{equation}
T^{\tilde{a}}=x^{\tilde{a}}{}_{,\mu}T^\mu
=x^{\tilde{a}}{}_{,i}T^i+x^{\tilde{a}}{}_{,4}T^{{}_4}
\end{equation}
which yields $T^{\tilde{a}}=x^{\tilde{a}}{}_{,i}T^i$ only in the absence of the 4th component, while the 4th component $T^{\tilde{4}}=x^{\tilde{4}}{}_{,4}T^{{}_4}=T^{{}_4}$ is identically 0 (the Latin indices $\tilde{a},i (=1,2,3)$ represents spatial components while the Greek $\mu (=1,2,3,4)$ for space-time component). In general, the tensor field with zero time components transforms as
\begin{equation}
T^{\tilde{a}\tilde{b}\cdots}{}_{\tilde{c}\tilde{d}\cdots}
=x^{\tilde{a}}{}_{,i}x^{\tilde{b}}{}_{,j}\cdots x^k{}_{,\tilde{c}}x^l{}_{,\tilde{d}}\cdots T^{ij\cdots}{}_{kl\cdots},
\label{3D transform}
\end{equation}
which behaves as a tensor in $\mathscr{S}_t$. Let us call tensor of this class the \emph{3-D tensor}. Eq. (\ref{3D transform}) means that any 3-D tensor remains 3-D tensor under the non-relativistic transformation (\ref{S transform}); namely (\ref{S transform}) forms the covariance group of 3-D tensor. Since we still remain on the non-relativistic formalism, we apply the term \emph{general covariance} to the diffeomorphism covariance of 3-D tensors under the transformation (\ref{S transform}) \citep{metric}. The general covariance of 3-D tensors and their theories are obvious as far as they are written in the differential geometry of 4-D space-time, which is the biggest merit of the space-time formalism.

\section{Mean-Lagrangian formalism} 
\subsection{Mean flow}

\begin{figure}
\centering
\includegraphics[width=8.5cm]{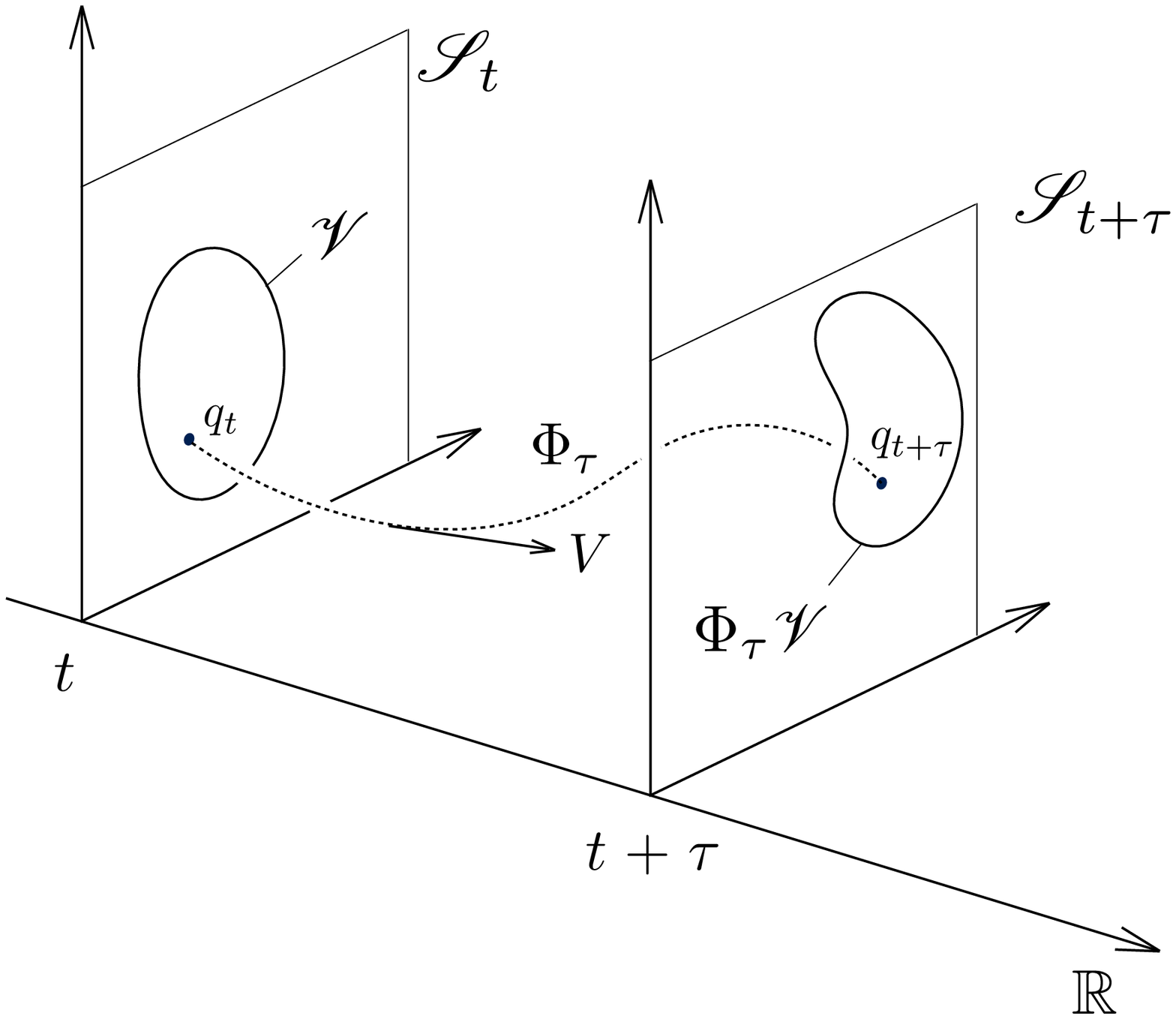}
\caption{Mean flow $\Phi_\tau$ is naturally introduced as a vector flow of a 4-dimensional vector $V=(\mathbf{V},1)$. An arbitrary open set $\mathscr{V}(\in \mathscr{S}_t)$ may be mapped to $\Phi_\tau \mathscr{V}$, which can be interpreted as the motion of virtual continuum. }
\label{mean flow}
\end{figure}

The fluid element, like a particle, is identified by its space-time trajectory $p_t=(\mathbf{p}(t),t)(\in\mathscr{M})$, where we denote the space component (projection on $\mathscr{S}_t$) by the bold type. The total fluid motion is identified by the flow (one-parameter group) $\phi_\tau:\mathscr{M}\times\mathbb{R}\rightarrow\mathscr{M}$ such that $\phi_\tau p_t=p_{t+\tau}$. Then a vector field $v: \mathscr{M}\rightarrow T\mathscr{M}$ is defined such that
\begin{equation}
v(\phi_\tau p_t)=\frac{d}{d\tau}\phi_\tau p_t
\end{equation}
for all existing $p_t$, whose coordinate representation reads $v=(v^{{}_1}(\mathbf{x},t),v^{{}_2}(\mathbf{x},t),v^{{}_3}(\mathbf{x},t),1)$, where $v^{{}_4}=1$. The projection $\mathbf{v}$ on $\mathscr{S}_t$, gives the velocity field observed in $\mathscr{S}_t$. Now the coordinate transformation reads
\begin{subequations}
\begin{align}
v^{\tilde{a}}&=x^{\tilde{a}}{}_{\,\mu}v^\mu=x^{\tilde{a}}{}_{,i}v^j+x^{\tilde{a}}{}_{,t}v^{{}_4}=x^{\tilde{a}}{}_{,i}v^j+x^{\tilde{a}}{}_{,t},\label{v rule}\\
v^{{}_{\tilde{4}}}&=x^{{}_{\tilde{4}}}{}_{\,\mu}v^\mu=\tilde{t}{}_{,i}v^j+\tilde{t}{}_{,t}v^{{}_0}=v^{{}_0}=1\label{v0 rule},
\end{align}
\end{subequations}
which is identical to what is derived in \citep{Ariki15a}.

In the above procedure we saw the derivation of the velocity field $\mathbf{v}(\mathbf{x},t)$ from the flow $\phi_\tau$, which may be summarized that $\phi_\tau$ is the vector flow uniquely determined by the vector field $v$. This one-to-one correspondence can be cast into the mean velocity; another vector flow, say $\Phi_\tau$, exists for the mean velocity field. As far as $\mathbf{v}(\mathbf{x},t)$ of each realization is smooth and single-valued, its ensemble average $\mathbf{V}(\mathbf{x},t)\equiv\langle v(\mathbf{x},t)\rangle$ is also smooth and single-valued. Then we compose another space-time vector field $V=(\mathbf{V},1):\mathscr{M}\rightarrow T\mathscr{M}$. Needless to say, the transformation rules (\ref{v rule}) and (\ref{v0 rule}) with replacement of $\mathbf{v}$ by $\mathbf{V}$ hold:
\begin{subequations}
\begin{align}
V^{\tilde{a}}&=x^{\tilde{a}}{}_{\,\mu}V^\mu=x^{\tilde{a}}{}_{,i}V^j+x^{\tilde{a}}{}_{,t}V^{{}_4}=x^{\tilde{a}}{}_{,i}V^j+x^{\tilde{a}}{}_{,t},\label{V rule}\\
V^{{}_{\tilde{4}}}&=x^{{}_{\tilde{4}}}{}_{\,\mu}V^\mu=\tilde{t}{}_{,i}V^j+\tilde{t}{}_{,t}V^{{}_0}=V^{{}_0}=1\label{V0 rule},
\end{align}
\end{subequations}
which are also directly obtained by averaging (\ref{v rule}) and (\ref{v0 rule}). According to the Frobenius theorem, a 1st-order differential equation 
\begin{equation}
\frac{d}{dt}q_t=V(q_t)
\label{q eq.}
\end{equation} 
always has a unique solution $q_t$ for a proper initial condition, which may be represented by a one-parameter group $\Phi_\tau: q_t\mapsto q_{t+\tau}$. By assembling each solution, we obtain another flow $\Phi_\tau:\mathscr{M}\times\mathbb{R}\rightarrow\mathscr{M}$, where an arbitrary open set $\mathscr{V}(\in \mathscr{S}_t)$ is transported to $\Phi_\tau \mathscr{V}(\in\mathscr{S}_{t+\tau})$ as time passes (figure \ref{mean flow}). Unlike the original flow $\phi_\tau$, there is no physically observable point convected by the mean flow, so $\Phi_\tau$ introduces rather a \emph{virtual} continuum. In spite of the absence of some observable continuum, we can construct the convective picture based on the mean flow without mathematical flaw. Once we obtain the mean flow $\Phi_\tau$, its resulting properties should be equal to what follows from $\phi_\tau$, which is the underlying truth of the $v$-$V$ correspondence of \citep{Ariki15a}. Finally we mention about the velocity fluctuation; the velocity fluctuation in 4-dimensional framework is given by $v'\equiv v-V=(\mathbf{v}',0)$, which is a 3-D vector field in $\mathscr{S}_t$ as also shown in \citep{Ariki15a}. 
\subsection{Mean-convective operations}
In general, tensor fields of different points cannot be treated equally, so their derivative and integration demand some mapping between different tensor spaces. In our formalism, we shall consider the difference in both space and time. The spatial difference, on one hand, can be bridged by the parallel transport based on the Levi-Civita connection. For the time difference, on the other hand, we need the pullback (or pushforward) of the flows $\phi_\tau$ or $\Phi_\tau$. Let us consider the general tensor field $T$ on the space-time $\mathscr{M}$. Although $T_{t+\tau}$ at time $t+\tau$ cannot treated as the tensor field at time $t$, $\phi_\tau^* T_{t+\tau}$ can be treated as the tensor field at time $t$. Using this pullback formalism, one can define the generally covariant derivative and integration. As an example, the derivative operation
\begin{equation}
\frac{\mathfrak{d}}{\mathfrak{d}t}T_t=\underset{\tau\rightarrow 0}{lim} \frac{\phi^*_\tau T_{t+\tau}-T_t}{\tau}=\mathcal{L}_v T_t
\label{conv der.}
\end{equation} 
gives a tensor field at time $t$. Note that this is actually the \emph{Lie derivative} $\mathcal{L}_v$ by the flow vector $v$. For a general 3-D tensor $T$, Eq. (\ref{conv der.}) yields 
\begin{equation}
\begin{split}
\frac{\mathfrak{d}}{\mathfrak{d}t}T^{ij\cdots}{}_{kl\cdots}
&=\frac{\partial}{\partial t} T^{ij\cdots}{}_{kl\cdots}+ T^{ij\cdots}{}_{kl\cdots;a} v^a\\
&\ \ \ \ -T^{aj\cdots}{}_{kl\cdots} {v^i}_{;a}-T^{ia\cdots}{}_{kl\cdots} {v^j}_{;a}-\cdots\\
&\ \ \ \ +T^{ij\cdots}{}_{bl\cdots} {v^b}_{;k}+T^{ij\cdots}{}_{kb\cdots} {v^b}_{;l}+\cdots.
\end{split}
\label{conv der. in S}
\end{equation}
which is exactly equivalent to the \emph{convective derivative}, measuring a physically objective tensor rate \citep{Old50}. The general covariance under (\ref{S transform}) is obvious from the 4-dimensional diffeomorphism covariance. Its inverse operation, the \emph{convective integration}, is defined by
\begin{equation}
\int^{t_1}_{t_2} T_t\mathfrak{d}\tau=\int^{t_1}_{t_2}\phi^*_\tau T_{t+\tau}\mathrm{d}\tau,
\label{conv int.}
\end{equation}
which gives the tensor field at time $t$ by accumulating the history of tensor $T$ within the time range $[t+t_2,t+t_1]$ along the flow path. Indeed $\int\mathfrak{d}t$ gives the integration of the tensor $T$ along the flow.

Extension of (\ref{conv der.}) and (\ref{conv int.}) to the $\Phi_\tau$-based operations is straightforward: corresponding to Eq. (\ref{conv der.}) we obtain 
\begin{equation}
\frac{\mathfrak{D}}{\mathfrak{D}t}T_t=\underset{\tau\rightarrow 0}{lim} \frac{\Phi^*_\tau T_{t+\tau}-T_t}{\tau}=\mathcal{L}_V T_t
\label{mconv der.}
\end{equation} 
which gives another definition of the \emph{mean-convective derivative} in \citep{Ariki15a}. The inverse operation becomes
\begin{equation}
\int^{t_1}_{t_2} T_t\mathfrak{D}\tau=\int^{t_1}_{t_2}\Phi^*_\tau T_{t+\tau}\mathrm{d}\tau,
\label{mconv int.}
\end{equation}
which may be referred as the \emph{mean-convective integration}, accumulating the history of tensor $T$ along the mean-flow path. Note that both operations (\ref{mconv der.}) and (\ref{mconv int.}) provide generally covariant operations for general 3-D tensors, which trivially follows from the original (\ref{conv der.}) and (\ref{conv int.}). Likewise, due to equivalence between $\phi_\tau$ and $\Phi_\tau$, the mean-Lagrangian framework offers objective operations in totally the same manner as the ordinary Lagrangian framework.
\subsection{Mean-Lagrangian coordinate frame}
As is wellknown, the convective operations (\ref{conv der.}) and (\ref{conv int.}) in a general-coordinate frame require complex calculations due to its pullback formalism. One can avoid such complexity by choosing the Lagrangian coordinate frame, which provides the strong tool in the actual calculations. Here we show that such coordinate representation is also introduced for the mean flow $\Phi_\tau$. 

We choose the coordinate representation so that the vector field $V$ becomes $(\mathbf{0},1)$. In this representation, $\Phi_\tau$ is only the time advancement while identity for the spatial component, where $\mathfrak{D/D}t$ and $\int\mathfrak{D}t$ become simply $\partial/\partial t$ and $\int\mathrm{d}t$. Such space-time coordinate frame, say $\{\mathbf{y},t\}$, can be obtained by using (\ref{V rule}):
\begin{equation*}
\begin{split}
0&=V^\mathfrak{a}(\mathbf{y},t)\\
&=\frac{\partial y^\mathfrak{a}}{\partial x^i}(\mathbf{x}(\mathbf{y},t),t)V^i(\mathbf{x}(\mathbf{y},t),t)+\frac{\partial y^\mathfrak{a}}{\partial t}(\mathbf{x}(\mathbf{y},t),t)\\
&=\frac{\partial y^\mathfrak{a}}{\partial x^i}(\mathbf{x}(\mathbf{y},t),t)\left\{V^i(\mathbf{x}(\mathbf{y},t),t)-\frac{\partial}{\partial t}x^i(\mathbf{y},t)\right\},
\end{split}
\end{equation*}   
hence
\begin{equation}
\frac{\partial}{\partial t}x^i(\mathbf{y},t)=V^i(\mathbf{x}(\mathbf{y},t),t).
\label{mconv cor.}
\end{equation}
This is actually equivalent to Eq. (\ref{q eq.}), and the solution $\mathbf{x}(\mathbf{y},t)$ exist from the Frobenius theorem. According to Eq. (\ref{mconv cor.}), the solution $\mathbf{x}(\mathbf{y},t)$ for fixed $\mathbf{y}$ represents the point transported by the mean velocity $\mathbf{V}$, so that $\{\mathbf{y},t\}$ offers the co-moving frame based on the mean flow $\Phi_\tau$, which may be referred as the \emph{mean-Lagrangian coordinate frame}. 

\section{Applications} 
\subsection{Visco-elastic effect}
Starting from the early works \citep{Lum70, Crow68}, the history effect and the corresponding visco-elastic features are pointed out as an important feature of general turbulence. Let us see that such visco-elastic feature can be represented in generally covariant manner, using a simple turbulence model as an example. As remarked in \citep{Ariki15a}, the known Launder-Reece-Rodi model \citep{LRR75} can be reformulated in a generally covariant form:
\begin{equation}
\begin{split}
\frac{\mathfrak{D}}{\mathfrak{D}t}R^{ij}
=&(1-C_{IP})(S^i_kR^{jk}+S^j_kR^{ik})\\
&+(1-C_{IP})(\Theta^i{}_kR^{jk}+\Theta^j{}_kR^{ik})\\
&-C_R\frac{\epsilon}{K}\left(R^{ij}-\frac{2}{3}Kg^{ij}\right)-\frac{2}{3}C_{IP}\mathsfbi{S\cdot R}\\
&-\frac{2}{3}\epsilon g^{ij},
\end{split}
\label{LRR}
\end{equation}
where the diffusion effects are neglected. The equation for the deviatoric stress $\mathsfbi{P}\equiv \frac{2}{3}K\mathsfbi{g}-\mathsfbi{R}$ becomes
\begin{equation}
\begin{split}
\left(\frac{\mathfrak{D}}{\mathfrak{D}t}+C_R\frac{\epsilon}{K}\right)P^{ij}
=\frac{2}{3}(1-2C_{IP})KS^{ij}+\cdots
\end{split}
\label{LRR 2}
\end{equation}
The operator $(\mathfrak{D}/\mathfrak{D}t-C_R\epsilon/K)$ is rewritten in the mean-Lagrangian coordinates as $(\partial/\partial t-C_R\epsilon/K)$ whose Green function is given by $\mathrm{exp}[-\int^t_{t'}\frac{K_\tau}{C_R\epsilon_\tau}\mathrm{d}\tau]$. Thus the inverse operation of $(\mathfrak{D}/\mathfrak{D}t-C_R\epsilon/K)$ is given in general by
\begin{equation*}
\begin{split}
\left(\frac{\mathfrak{D}}{\mathfrak{D} t}+C_R\frac{\epsilon}{K}\right)^{-1}
= \int^t\mathfrak{D}t'\mathrm{exp}
\left[-\int^t_{t'}\frac{K_\tau}{C_R\epsilon_\tau}\mathfrak{D}\tau\right]\times
\end{split}
\label{inverse}
\end{equation*}
Thus the eddy-viscosity term (the first term of the right-hand side) of (\ref{LRR 2}) can be integrated as
\begin{equation}
\begin{split}
P_t^{ij}=\frac{2}{3}(1-2C_{IP})\int^t_{-\infty}&\mathfrak{D}t'\mathrm{exp}
\left[-\int^t_{t'}\frac{K_\tau}{C_R\epsilon_\tau}\mathfrak{D}\tau\right]\\
&\times K_{t'}S^{ij}_{t'}+\cdots,
\end{split}
\label{viscoelastic}
\end{equation}
which exactly represents the visco-elastic effect in generally covariant form; namely, (\ref{viscoelastic}) represents physically objective properties free from coordinate representation. If one employs, on the contrary, the non-covariant equation such as 
\begin{equation}
\begin{split}
\frac{\mathrm{D}}{\mathrm{D}t}R^{ij}=&(1/2-C_{IP})(S^i_kR^{jk}+S^j_kR^{ik})\\
&(1/2-C_{IP})(\Theta^i{}_kR^{jk}+\Theta^j{}_kR^{ik})+\cdots,
\end{split}
\label{broken LRR}
\end{equation}
one may obtain 
\begin{equation}
\begin{split}
\!\!P_t^{ij}=\frac{2}{3}\left(\frac{1}{2}-2C_{IP}\right)\int^t_{-\infty}&\mathrm{D}t'\mathrm{exp}
\left[-\int^t_{t'}\frac{K_\tau}{C_R\epsilon_\tau}\mathrm{D}\tau\right]\\
&\times K_{t'}S^{ij}_{t'}+\cdots,
\end{split}
\label{broken viscoelastic}
\end{equation}
instead of Eq. (\ref{viscoelastic}). Here $\int\mathrm{D}t$ is the inverse of $\mathrm{D/D}t$ which is also employed in \citep{Crow68} to obtain the visco-elastic feature of the Reynolds stress. $\int\mathrm{D}t$ incorporates only the translation but not the distortion by the mean flow, so Eq. (\ref{broken viscoelastic}) fully depends on the coordinate representation, loosing its physical objectivity. Finally, it is notable that the above analysis consists for any models containing the Rotta term, not only for the LRR model (\ref{LRR}).

\subsection{Extended Lagrangian correlations}

The main stream of the multiple-points closure have been developed for the two-point quantities, where two-point correlations obey differential-integral equations (the time integration are sometime simplified by the Markovianization). Here we should remark that the simple time derivative and integration depend on the coordinate representation, and their physical interpretation may change as the reference frame does. The Lagrangian (mean-Lagrangian) description allows us to avoid this difficulty by incorporating the convective (mean-convective) operations, guaranteeing covariant formulation of the closure theories of various types. 

Let us compose covariant correlation between $v'_x$ and $v'_{x'}$ at different space-time points $x=(\mathbf{x},t)$ and $x'=(\mathbf{x}',t')$, where $v'_x(\in{T_x \mathscr{S}_t})$ is to be mapped onto $T_{x'}\mathscr{S}_{t'}$ so that we obtain a correlation tensor in $T_{x'}\mathscr{S}_{t'}\otimes T_{x'}\mathscr{S}_{t'}$. When we compare $v'$ at different times, as remarked in the previous discussions, we need either of $\phi_\tau^*$ and $\Phi_\tau^*$ to keep the general covariance ($\tau=t-t'$). Let us first apply $\phi_\tau^*$, where $v'_x$ is mapped to $\phi_\tau^*v'_x(\in T_{\phi^{-1}_\tau x}\mathscr{S}_{t'})$ at time $t'$. Then we still have spatial gap between $\phi^{-1}_\tau x$ and $x'$, which may be bridged by the parallel transport  $P_{\phi^{-1}_\tau x}^{x'}$ defined by the Levi-Civita connection of $\mathscr{S}_{t'}$, which does not depend on its path in the flat Riemann space. The vector $v'_x$ is mapped in the following steps:
\begin{equation}
T_x\mathscr{S}_t\overset{\phi^*_\tau}{\longrightarrow} T_{\phi^{-1}_\tau x}\mathscr{S}_{t'}
\overset{P}{\longrightarrow}T_{x'}\mathscr{S}_{t'}.
\end{equation} 
Then we obtain
\begin{equation}
{}^{{}_F}\!U(x,x')\equiv\langle P_{\phi^{-1}_\tau x}^{x'}(\phi_\tau^* v'_x)
\otimes v'_{x'}\rangle. 
\label{fine-Lagrangian correlation}
\end{equation}
To make a clear difference from the mean-Lagrangian picture, let us call ${}^{{}_F}\!U$ as the \emph{fine-Lagrangian correlation}, which reflects the fine-motion of the fluid element. Although this is similar to so called \emph{Lagrangian velocity correlation} (${}^{{}_L}\!U$) which also considers the translation of fluid element by its turbulence motion, there is a crucial difference; in addition to the translation, Eq. (\ref{fine-Lagrangian correlation}) explains also the distortion of the fluid element due to pullback $\phi^*_\tau$. Another quantity is obtained by replacing $\phi_\tau$ with $\Phi_\tau$: 
\begin{equation}
{}^{{}_M}\!U(x,x')\equiv\langle P_{\Phi^{-1}_\tau x}^{x'}(\Phi_\tau^* v'_x)
\otimes v'_{x'}\rangle,
\label{mean-Lagrangian correlation}
\end{equation}
which may be called as the \emph{mean-Lagrangian correlation}, incorporating the convection effect by the mean flow. However, by its definition, ${}^{{}_M}\!U$ does not consider the convection of the detailed motion of the fluid element, so it fails to extract the fine-scale properties of K41 \citep{K41a,K41b}, as is the same situation for the Eulerian correlation (${}^{{}_E}\!U$). Indeed, ${}^{{}_M}\!U$ coincides with the ${}^{{}_E}\!U$ in the absence of the mean velocity field $\mathbf{V}$. 

${}^{{}_F}\!U$, in contrast to the others (${}^{{}_L}\!U$, ${}^{{}_E}\!U$, and ${}^{{}_M}\!U$), is the only correlation that considers the fine-scale properties in generally-covariant manner, so ${}^{{}_F}\!U$ could be employed to measure the objective properties of turbulence along with the Kolmogorov scaling \citep{K41a,K41b}. One shortcoming arises, however, when we apply the perturbation analysis on the basis of \citep{Krai65,Kaneda81,KG97,Krai59,Wyld61}; the pullback $\phi_\tau^*$ considers the distortion of the fluid element that causes further nonlinearity in $v'$; namely ${}^{{}_F}\!U$ sacrifices its mathematical simplicity for the general covariance. In the actual analysis, we may need some other quantities with enough feasibility. One possible solution is to modify ${}^{{}_M}\!U$; we incorporate the translation of fluid element by replacing $x$ with $\phi_\tau \bar{x}'$ ($\bar{x}'\in \mathscr{S}_{t'}$), namely   
\begin{equation}
\begin{split}
\!\!{}^{{}_D}\!U(\bar{x}',x';\tau)&\equiv {}^{{}_M}\!U(\phi_\tau \bar{x}',x')\\
&=\langle P_{\Phi^{-1}_\tau\circ \phi_\tau \bar{x}'}^{x'} (\Phi_\tau^* v'_{\phi_\tau \bar{x}'}) \otimes v'_{x'}\rangle,
\end{split}
\label{double-Lagrangian correlation}
\end{equation}
where both $\phi_\tau$ and $\Phi_\tau$ appear. Let us call (\ref{double-Lagrangian correlation}) as the \emph{double-Lagrangian correlation} for its mixed scheme of fine- and mean-Lagrangian frameworks.

\begin{figure}
\centering
\includegraphics[width=7.5cm]{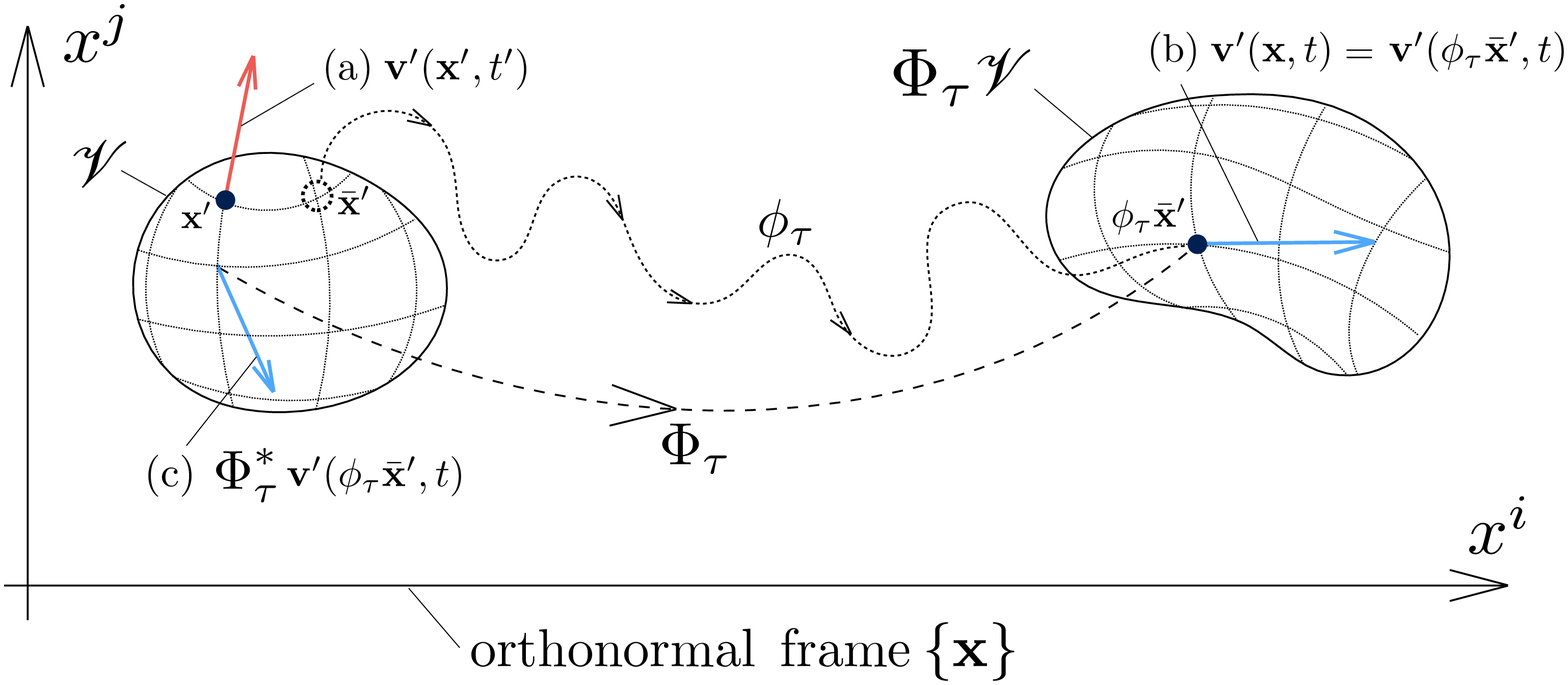}
\caption{The difference between ${}^{{}_D}\!U$ and ${}^{{}_L}\!U$ is schematically explained in 3-D space with orthonormal coordinate setup $\{\mathbf{x}\}$. The Lagrangian correlation ${}^{{}_L}\!U$ is given by $\langle$(a)$\otimes$(b)$\rangle$, while the double Lagrangian correlation ${}^{{}_D}\!U$ is $\langle$(a)$\otimes$(c)$\rangle$. Namely, in the double-Lagrangian correlation, the velocity fluctuation is measured in the material frame of the virtual-continuum bulk $\mathscr{V}$ (the mean-Lagrangian coordinate frame).}
\label{comparison}
\end{figure}

It may be instructive to show the difference between the Lagrangian ${}^{{}_L}\!U$ and double-Lagrangian ${}^{{}_D}\!U$ in purely spatial description, which is sketched in figure \ref{comparison}. For simplicity, we consider the orthonormal frame for spatial coordinate, so the parallel transport $P$ becomes trivial. We consider the correlation between (a) $\mathbf{v}'(\mathbf{x}',t')$ and (b) $\mathbf{v}'(\mathbf{x},t)$. Detailed fluid motion is considered by replacing $\mathbf{x}$ of (b) with $\phi_\tau \bar{\mathbf{x}}'$. Then ${}^{{}_L}\!U$ gives the correlation between (a) and (b). In the double-Lagrangian formulation, on the contrary, the fluctuation vector (b) is mapped to (c) by the pullback $\Phi_\tau^*$; namely, $\mathbf{v}'(\phi_\tau\bar{\mathbf{x}}',t)$ should be measured by the comoving frame of the virtual-continuum bulk ($\mathscr{V}$ in figure \ref{comparison}) convected by the mean flow, which implies that the mean-Lagrangian coordinate frame gives the simplest expression. Unlike (c), (b) is measured in a different way under the coordinate transformation (\ref{S transform}). The double-Lagrangian correlation ${}^{{}_D}\!U$ is generally covariant under the non-relativistic general transformation (\ref{S transform}), while the Lagrangian correlation ${}^{{}_L}\!U$ is limited to the Galilean covariance. Thus the double-Lagrangian correlation is a natural extension of the Lagrangian correlation with far wider covariance group.

For simplicity of calculations, let us consider a limited class of coordinate system $\{z^{{}_1},z^{{}_2},z^{{}_3},z^{{}_4}\}=\{\mathbf{z},t\}$
where the metric becomes constant in space (geodesic coordinate frame). This forms a Lie group of time-dependent affine transformation: 
\begin{equation}
z^{A^*}=X^{A^*}{}_{I}(t)z^I+Y^{A^*}(t)\, (A^*,I=1,2,3),
\label{affine}
\end{equation}
where $X^{A^*}{}_{I}(t)$ and $Y^{A^*}(t)$ are both independent of coordinates. In this coordinate representation, the parallel transport $P$ in Eq. (\ref{double-Lagrangian correlation}) becomes identity $\mathbf{1}$, so that we have the following simpler form:
\begin{equation}
\!\!{}^{{}_D}\!U(\bar{z}',z';\tau)\equiv \langle \Phi_\tau^* v'_{\phi_\tau \bar{z}'} \otimes v'_{z'}\rangle.
\label{simple DL}
\end{equation}
The velocity fluctuation obeys
\begin{equation}
\begin{split}
\!\!\!\!\left(\frac{\mathfrak{D}}{\mathfrak{D} t}-\nu\triangle \right)\!v'^I\!+v'^I{}_{,J}v'^J\!+p'^{,I}
=\!-\!\left(S^I_J+\Theta^I{}_J\right)\!v'^J,
\end{split}
\label{v' eq.}
\end{equation} 
and incompressibility $v'^J{}_{,J}=0$, where $\mathsfbi{S}$ and $\mathsfbi{\Theta}$ are the \emph{strain rate} and \emph{absolute vorticity} of the mean flow \citep{Ariki15a}. $v'_{\phi_\tau \bar{z}'}$ can be represented by the \emph{Lagrangian-position function} $\psi$ \citep{Krai65,Kaneda81}:
\begin{equation}
\left\{\frac{\mathfrak{D}}{\mathfrak{D} t}+v'^J(\mathbf{z}'',t)\partial''_J\right\}\psi(\mathbf{z}'',t;\bar{\mathbf{z}}',t')=0,
\label{psi eq.}
\end{equation}
then we have
\begin{equation*}
\begin{split}
v'_{\phi_\tau \bar{z}'}
\rightarrow\int \sqrt{\mathrm{det}[g_{IJ}(t)]}\mathrm{d}^3z''\psi(\mathbf{z}'',t;\bar{\mathbf{z}}',t')v'^I(\mathbf{z}'',t).
\end{split}
\end{equation*}
Eqs. (\ref{v' eq.}) and (\ref{psi eq.}) form similar set of equations to that of \citep{Krai65,Kaneda81}, but the convective derivative $\mathfrak{D/D}t$ appears in stead of simple time derivative $\partial/\partial t$. Then ${}^{{}_D}\!U$ becomes
\begin{equation}
\begin{split}
&{}^{{}_D}\!U^{IJ}(\bar{z}',z';\tau)\\
&=\int \sqrt{\mathrm{det}[g_{IJ}(t)]}\mathrm{d}^3z''\langle\psi(\mathbf{z}'',t;\bar{\mathbf{z}}',t')v'^I(\mathbf{z}'',t)v'^J(\mathbf{z}',t')\rangle.
\end{split}
\label{UD in z}
\end{equation}
Using Eqs. (\ref{v' eq.}) and (\ref{psi eq.}), one can obtain the closure equations containing the mean-convective operations $\mathfrak{D/D}t$ and $\int\mathfrak{D}t$, which may describe the physically objective dynamics of (\ref{UD in z}).

\section{Conclusion}
In this paper, we discussed about the theoretical basis of the mean-Lagrangian framework in the fluid turbulence, which provides some convenient mathematical tools such as the mean-convective derivative and integration, and the mean-Lagrangian coordinate frame as natural consequences. The mean-Lagrangian framework applied to a one-point closure model enables the explicit integration of the visco-elastic effect on the Reynolds stress. In the two-point closure theories, we proposed some new turbulence quantities such as ${}^{{}_F}\!U$, ${}^{{}_M}\!U$, and ${}^{{}_D}\!U$ allowed by the general covariance. Among them, ${}^{{}_D}\!U$, the double-Lagrangian correlation, gives the natural generalization of the ordinary Lagrangian correlation of \citep{Krai65,Kaneda81} by expanding its covariance group from simple Galilean to the general non-relativistic transformation group of (\ref{S transform}). In this paper, we do not go further steps by choosing some particular closure scheme, since the validity of all the discussions given so far is totally scheme-independent. To obtain the total performance of the resulting model, we should construct the closure equations, at least, for (\ref{UD in z}) on the basis of Eqs. (\ref{v' eq.}) and (\ref{psi eq.}), which will be reported in the forthcoming paper \citep{perspective}. The first step for such closure modeling may be conducted for the homogeneous anisotropic turbulence, where further simplification can be made by using the mean-Lagrangian coordinate frame which is briefly discussed in Appendix \ref{HAT}.

Thus it is notable that, although the mean-Lagrangian framework offers a suitable picture that automatically incorporates the physical objectivity via the general covariance, it does not necessarily promise the quantitative successes. As widely recognized, various aspects to be clarified have been left behind their approximation procedures, which prevents us from reaching the consensual physical modeling, if exists. Nevertheless, the covariance principle should be firmly considered since it provides an undisputed basis of physics that is to describe the objective properties of nature, irrespective of the choice of modeling schemes. In such covariant formulation of the turbulence theory, the mean-Lagrangian framework offers universal strategies applicable to wide range of schemes and phenomena.

\subsection*{Acknowledgments}
The present work is supported by the Research Fellowships of Japan Society for the Promotion of Science (JSPS), and also partially by JSPS KAKENHI Grant No. (S) 24224003.



\appendix

\section{Homogeneous anisotropic turbulence}\label{HAT}

In case of the homogeneous anisotropic turbulence, one can choose the mean-Lagrangian frame $\{y\}=\{\mathbf{y},t\}$ for $\{z\}=\{\mathbf{z},t\}$, where Eqs. (\ref{v' eq.}) and (\ref{psi eq.}) are reduced to
\begin{equation}
\begin{split}
\!\!\!\!\left(\frac{\partial}{\partial t}-\nu\triangle \right)\!v'^{\mathfrak{a}}+v'^{\mathfrak{a}}{}_{,\mathfrak{b}}v'^{\mathfrak{b}}+p'^{,\mathfrak{a}}
=\!-\!\left(S^\mathfrak{a}_\mathfrak{b}+\Theta^\mathfrak{a}{}_\mathfrak{b}\right)\!v'^{\mathfrak{b}},
\end{split}
\label{v' eq. in y}
\end{equation} 
\begin{equation}
\left\{\frac{\partial}{\partial t}
+v'^{\mathfrak{a}}(\mathbf{y}'',t')\partial''_\mathfrak{a}\right\}\psi(\mathbf{y}'',t';\bar{\mathbf{y}}',t)=0,
\label{psi eq. in y}
\end{equation}
which effectively reduces the complexity in the calculus of the mean-convective derivative $\mathfrak{D/D}t$. Also one should add the time evolution of the metric tensor by 
\begin{equation}
\frac{d}{d t} g_\mathfrak{ab}(t)=S_\mathfrak{ab}(t),
\label{g eq. in y}
\end{equation}
where both $g_\mathfrak{ab}(t)$ and $S_\mathfrak{ab}(t)$ do not depends on spatial position. Then Eq. (\ref{UD in z}) becomes
\begin{equation}
\begin{split}
&{}^{{}_D}\!U^\mathfrak{ab}(\bar{y}',y';\tau)\\
&=\int\mathrm{d}^3y''\left\langle \psi(\mathbf{y}'',t';\bar{\mathbf{y}}',t) v'^{\mathfrak{a}}(\mathbf{y}'',t')v'^{\mathfrak{b}}(\mathbf{y}',t)\right\rangle.
\end{split}
\end{equation}
On the basis of Eqs. (\ref{v' eq. in y}), (\ref{psi eq. in y}), and (\ref{g eq. in y}), one can develop the closure models following similar strategies developed by pioneers \citep{Krai65,Kaneda81}. One should remark, however, that the time derivative $\partial/\partial t$ and the time integration $\int\mathrm{d}t$ in such closure models are actually the mean-convective derivative $\mathfrak{D/D}t$ and integration $\int\mathfrak{D}t$, which express the tensor rate and history effect in physically objective manner.


\begin{thebibliography}{99}
\bibitem{metric}
One may think the metric tensor does not satisfy this feature, since the metric in the relativistic formalism does have non-zero time components. However, the transformation (\ref{S transform}) does not preserve the Lorentz metric, and hence the four-dimensional metric structure does not exist. Thus the geometrical structure only exists in the spatial component, so we regard the metric as a 3-D tensor.
\bibitem{perspective}
Applications of \citep{Kaneda81} may be straightforward, which reduces to similar set of equations, the fluctuation-dissipation relation between the correlation and the response may be lost though. On the contrary, \citep{Krai59} and later \citep{KG97} do not seem trivial since the 3 wavenumbers of direct-interaction need to be labeled in a covariant manner during the time evolution of the frame, which may be not unique.
\end{thebibliography}
\end{document}